\documentclass[12pt,3p,review,authoryear]{elsarticle}

\usepackage{amsmath}
\usepackage{amssymb}
\usepackage[nolists,nomarkers]{endfloat}
\usepackage[hyphens]{url}

\newcommand\blfootnote[1]{%
  {
  \renewcommand\thefootnote{}\footnote{#1}%
  \addtocounter{footnote}{-1}%
  }
}

\usepackage{colortbl}

\usepackage{xspace}

\usepackage{todonotes}
\newcommand{\mrev}[2]{#2}
\newcommand{\mrevi}[2]{#2}

\newcommand{\nrev}[2]{#2}

\newcommand{\refsupplphantomas}{S1}
\newcommand{\refsupplodfsim}{S2}

\usepackage{lineno}

\newcommand{\degree}{\ensuremath{^\circ}}

\newcommand{\vect}[1]{\ensuremath{\vec{#1}}}

\begin{document}

\title{{Fingerprinting Orientation Distribution Functions in Diffusion MRI detects smaller crossing angles}}

\author[CAIR,CBI]{Steven H. Baete\corref{cor1}}
\ead{steven.baete@nylangone.org}
\author[CAIR,CBI,Sackler]{Martijn A. Cloos}
\author[CAIR,CBI]{Ying-Chia Lin}
\author[Neuro]{Dimitris G. Placantonakis}
\author[CAIR,CBI]{Timothy Shepherd}
\author[CAIR,CBI]{Fernando E. Boada}
\address[CAIR]{Center for Advanced Imaging Innovation and Research (CAI$^2$R), NYU School of Medicine, New York, NY, USA}
\address[CBI]{Center for Biomedical Imaging, Dept. of Radiology, NYU School of Medicine, New York, NY, USA}
\address[Sackler]{The Sackler Institute of Graduate Biomedical Sciences, NYU School of Medicine, New York, NY, USA}
\address[Neuro]{Dept. of Neurosurgery, Perlmutter Cancer Center, Neuroscience Institute, Kimmel Center for Stem Cell Biology, NYU School of Medicine, New York, NY, USA}
\cortext[cor1]{Corresponding author at: NYU School of Medicine, Dept. of
Radiology, 660 1st Avenue, 4th Floor, New York, NY 10016, USA.}

\journal{NeuroImage}

\begin{abstract} 
Diffusion tractography is routinely used to study white matter architecture and brain connectivity in vivo. A key step for successful tractography of neuronal tracts is the correct identification of tract directions in each voxel. Here we propose a fingerprinting-based methodology to identify these fiber directions in Orientation Distribution Functions, dubbed ODF-Fingerprinting (ODF-FP).

In ODF-FP, fiber configurations are selected based on the similarity between measured ODFs and elements in a pre-computed library. In noisy ODFs, the library matching algorithm penalizes the more complex fiber configurations.

ODF simulations and analysis of bootstrapped partial and whole-brain \textit{in vivo} datasets show that the ODF-FP approach improves the detection of fiber pairs with small crossing angles while maintaining fiber direction precision, which leads to better tractography results.

Rather than focusing on the ODF maxima, the ODF-FP approach uses the whole ODF shape to infer fiber directions to improve the detection of fiber bundles with small crossing angle. The resulting fiber directions aid tractography algorithms in accurately displaying neuronal tracts and calculating brain connectivity.


\end{abstract}

\maketitle

\blfootnote{\noindent{\it Abbreviations\/}:
ADC, Apparent Diffusion Coefficient;
AIC, Akaike Information Criterion;
CHARMED, Composite Hindered and Restricted Diffusion Model;
CI, 95\% Confidence Intervals;
CSF, CerebroSpinal Fluid;
dODF, diffusion Orientation Distribution Function;
DSI, Diffusion Spectrum Imaging;
DTI, Diffusion Tensor Imaging;
DWI, Diffusion Weighted Imaging;
EPI, Echo Planar Imaging;
FA, Fractional Anisotropy;
fODF, fiber Orientation Distribution Function;
GQI, Generalized Q-Space Sampling;
HARDI, High Angular Resolution Diffusion Imaging;
HCP, Human Connectome Protocol;
ODF, Orientation Distribution Function;
ODF-FP, Orientation Distribution Function Fingerprinting;
QA, Quantitative Anisotropy;
qODF, q-ball Orientation Distribution Function;
RDSI, Radial Diffusion Spectrum Imaging;
ROI, Region of Interest;
SNR, Signal to Noise Ratio;
}

\section{Introduction}


Diffusion weighted MRI (DWI, \citep{LeBihan1986}) non-invasively captures the complex microstructure of the brain. The angular dependence of DWIs sensitivity to water molecule motion powers techniques such as Diffusion Tensor Imaging (DTI, \citep{Basser1996,Mori2002}), Q-ball imaging \citep{Tuch2002,Tuch2004} and Diffusion Spectrum Imaging (DSI, \citep{Wedeen2005,Wedeen2008,Baete2015RDSI,Baete2017}) to identify fiber bundle directions in each voxel. This fiber bundle information, collected throughout the whole brain, is the input for tractography algorithms which produce representations of long-range axonal structure \citep{Pierpaoli1996,Basser2000,Bammer2005,Fernandez-Miranda2012,Shin2012}. While the anatomical accuracy of these representations is a subject of much investigation \citep{Knosche2015,Schilling2016}, the ability to non-destructively obtain brain structural connectivity information has led to the adoption of tractography for use in neuroscience and clinical applications \citep{Fernandez-Miranda2012,Shin2012,Jbabdi2015,Galantucci2016,Mitra2016}.\par

High Angular Resolution Diffusion Imaging (HARDI) methods, such as multi-shell Q-ball and DSI, capture the complex intra-voxel crossings \citep{Wedeen2012,Fernandez-Miranda2012} in Orientation Distribution Functions (ODFs). Direct calculation of these ODFs (diffusion ODF or dODF) requires a sufficiently dense sampling of diffusion weightings and directions as in Cartesian \citep{Wedeen2005} and Radial DSI \citep{Baete2015RDSI}, typically distributed on several shells. From single shell acquisitions, q-ball ODFs (qODF) can be estimated using a spherical Radon transform \citep{Tuch2004}. Both these ODFs can be transformed to fiber ODFs (fODF) by spherical deconvolution with an estimated Fiber Response Function \citep{Tournier2008,Jeurissen2014,Dhollander2016} - the prototypical expected response of a single fiber.\par

A key step in employing ODFs in tractography algorithms, whether they are dODFs, qODFs or fODFs, is the correct identification of fiber directions in each voxel. To this end, many algorithms have been proposed ranging from simple finite difference
methods \citep{Descoteaux2007,Frey2008} to more complex numerical optimization solutions combining gradient ascent \citep{Berman2008} or Newton-Rhapson \citep{Tournier2004} algorithms with appropriate thresholds on fiber proximity. Other methods transform ODFs to a constrained polynomial basis to aid the numerical identification of stationary points on the ODFs surface \citep{Aganj2010}. Yet another approach uses Bayesian estimation to fit models to the diffusion data in each voxel (probabilistic estimation, \citep{Behrens2007}).\par

All these fiber direction identification approaches are however flawed due to the intrinsic ODF peak width (\citep{Barnett2009,Jensen2016}, Fig. \ref{odffpscheme}a). This ODF peak width \citep{Barnett2009} and the limited angular resolution of the acquisition make it difficult to accurately estimate the directions of fibers crossing at shallow angles (\citep{Kuo2008,Jeurissen2012}, Fig. \ref{odffpscheme}b). Most proposed methods indeed fail to detect crossing angles less than 40\degree \citep{Kuo2008,Jeurissen2012,Tournier2008,Descoteaux2007,Daducci2014,Wilkins2015}. Even deconvolving the ODFs with a Fiber Response Function fails to reliably detect crossing angles smaller than 30\degree \citep{Tournier2008,Jeurissen2014}. Failing to identify all bundle directions prohibits tractography algorithms from correctly following fibers in areas with crossing bundles.\par

By focusing on the maxima of the ODF to identify fiber directions, the methods listed above ignore the information captured in the shape of the ODF. For example, two fiber bundles crossing at an angle smaller than the intrinsic ODF peak width of the reconstruction give rise to one single ODF maximum. The flattened shape of the peak, however, reveals the presence of two fibers (Fig. \ref{odffpscheme}b).\par

Here we propose a new approach to fiber bundle identification inspired by key concepts first introduced in MR Fingerprinting \citep{Ma2013,Cloos2016}. Instead of a dictionary with spin evolutions at different $T_1$ and $T_2$ relaxation times, we generate a library of ODF-fingerprints and identify the fiber directions of ODFs by assessing the similarity between the measured data and the elements in our library (Fig \ref{odffpscheme}e). This ODF-Fingerprinting (ODF-FP) approach utilizes the whole ODF shape to infer fiber directions rather than more narrowly focusing on the ODF maxima. \par

In this work, we show that the ODF-FP method not only identifies smaller crossing angles more accurately, it also improves the performance at larger crossing angles. To this end, we discuss the different facets of the ODF-FP algorithm for fiber direction identification and look at the methods' performance in a simulated phantom containing crossing fiber bundles, at angular precision in individual simulated crossing fiber ODFs, \mrev{R2.2}{at the performance in a multi-resolution Human Connectome Protocol (HCP) dataset}, at reproducibility and noise sensitivity in \textit{in vivo} bootstrapped datasets and at the impact on tractography results in \textit{in vivo} whole brain RDSI acquisitions.\par

\section{Methods}

\subsection{ODF-Fingerprinting}

\begin{figure*}[tbh]
    \begin{center}
    \includegraphics[width=0.60\textwidth,trim=0 0 0 0, clip]{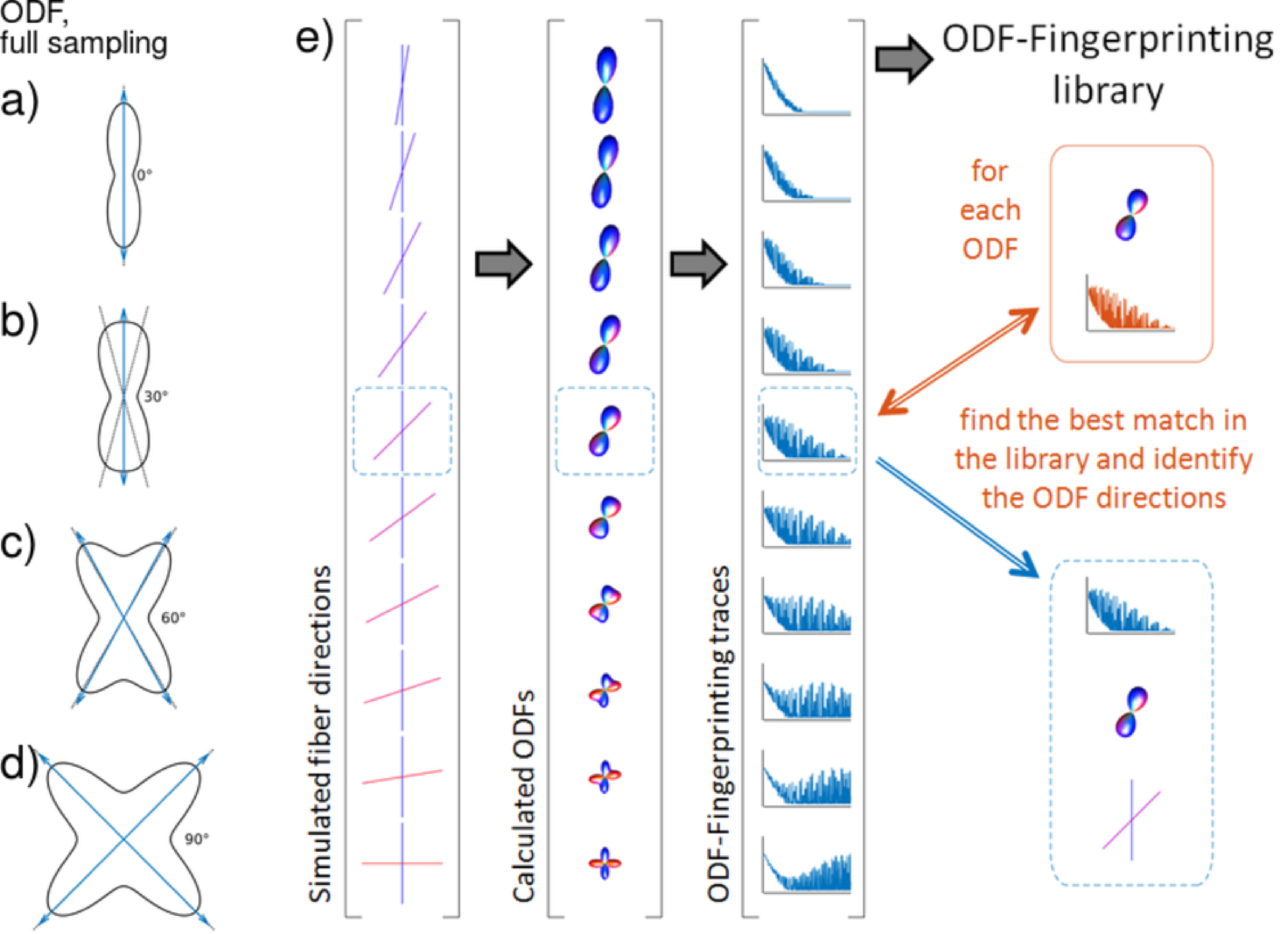}
        \caption{a-d) Exact ODFs of single (a) and crossing (b-d) fibers (FA$\,$=$\,$0.7) acquired with full q-space sampling. e) ODF-Fingerprinting: For each ODF of which the directions are to be determined, the proposed method searches the pre-constructed ODF library for the best match. Once the match is identified, the ODF's directions can be pulled from the library. \label{odffpscheme}}
    \end{center}
\end{figure*}

In ODF fingerprinting, measured ODFs are matched to a library of ODF-elements (Fig. \ref{odffpscheme}e). This library is generated by simulating diffusion weighted signals for a wide range of possible fiber combinations. Here we modeled each voxel as a composition of a water component $f_W$ and $N$ fibers, each with a volume fraction $f_j$ and a cylindrical diffusion tensor $\bar{D_j}$ ($\lambda_2\,=\,\lambda_3$). The simulated diffusion weighted signals are calculated as \citep{Alexander2001,Tuch2002,Wilkins2015}
\begin{align}
  S(\vect{q})\,=\,f_W e^{-\vect{q^2} D_W (\Delta-\frac{\delta}{3})} \label{ODFgen}\\
   +\,\Sigma_{j=1}^{N} f_j e^{-\vect{q}^T \bar{D}_j  \vect{q}(\Delta-\frac{\delta}{3})} \nonumber
\end{align}
with $\Delta$ and $\delta$ the diffusion time and the gradient duration, $D_W$ the water component diffusion coefficient and $\vect{q}$ the q-space wave vector. For each fiber combination, the simulated diffusion weighted signals are then reconstructed to form ODFs. These ODFs, expressed as vectors of numerical values are the entries of the ODF-library.\par

Matching of a measured $ODF_m$ is done by searching the ODF-dictionary for the ODF-entry with the best agreement. That is, for each $ODF_m$, we can find the index $l_m$ in the library matrix $L_{ODF}$ of the ODF-entry with the largest dot-product \citep{Ma2013,Cloos2016}:
\begin{equation}
  l_m\,=\,arg_l max(L_{ODF}\cdot ODF_m^T) \label{dotproduct}.
\end{equation}
Although this straightforward library matching algorithm works, we found that it tends to favor multiple fiber configurations in noisy ODFs \citep{Baete2018HBM}. Therefore, a penalty term was added to Eq. \ref{dotproduct}, which weighs complex fiber configurations in the library by a factor proportional to the noise estimate $\sigma_{n}$ of the input diffusion data, thus adapting the penalty to the dataset SNR. This approach is similar to model selection approaches such as the Akaike Information Criterion (AIC). The library matching algorithm then becomes
\begin{equation}
  l_m\,=\,arg_l max \left( log(L_{ODF}\cdot ODF_m^T) - \frac{n_{par}}{4 n} \sigma_{n} \right) \label{dotproductpenalty}
\end{equation}
with $n_{par}$ a measure of library element complexity, here $n_{par}\,=\,1 + 5 N$ ($N$ the number of fibers). $\sigma_{n}$ is estimated using a linear minimum mean square error estimator for the variance of the noise of the diffusion weighted signals \citep{Aja-Fernandez2008B}.\par

The size of the library is reduced by rotating the maximum values of both the library elements and $ODF_m$ to the Z-axis before matching. \mrev{R2.4}{Note that the ODF fingerprinting approach differs from earlier methods where dictionaries of ODF elements were learned from the acquired dataset using iterative compressed sensing methods \citep{Merlet2013,Bilgic2012,Gupta2017,Sun2013,Ye2012,Awate2013} and individual ODFs are reconstructed as a combination of a sparse subset of these learned elements. In contrast, in ODF-FP the library elements are generated independently of the acquired dataset based on a simple diffusion model (Eq. \ref{ODFgen}). The use of the diffusion model also distinguishes ODF-FP from methods where the library elements are generated from Monte-Carlo simulations of diffusing water in specific microstructural configurations \citep{Rensonnet2019}. }\par

The ODF-libraries used in this work simulate fiber combinations with up to \mrev{R1.1,R1.2}{3} fibers. The main fiber is aligned along the Z-axis and subsequent fibers are sampled on a 642-point tessellation of the unit sphere (angular resolution of 2.3\degree). Other parameters are: fiber FA ranging from 0.3 to 1.0 (in vivo) and 0.2 to 0.6 (simulations) in steps of 0.1 \mrev{R1.1}{(2 fibers; 0.16 for 3 fiber libraries)}; fiber bundle volume ranging from 0 to (100\%-water component) in steps of 10\% \mrev{R1.1}{(2 fibers; 0.16 for 3 fiber libraries)}; 10\% water component with an ADC of 0.9$\,$mm$^2$/s (in vivo) and 1.0$\,$mm$^2$/s (simulations)\mrev{R1.1}{; minimum angle between fibers is 20\degree in the 3 fiber libraries}. For the in-house datasets, diffusion signals are sampled on a Radial Diffusion Spectrum Imaging (RDSI, \citep{Baete2015RDSI}) grid with 236 q-space samples on four shells (250, 1000, 2250, 4000$\,$s$/$mm$^2$ (simulations) and 200, 1500, 2750, 4000$\,$s$/$mm$^2$ (in vivo)) and ODFs are generated with the RDSI-reconstruction. \mrev{R2.2}{In the HCP dataset, diffusion signals are sampled on 256 q-space points on three shells (1000, 3000, 5000$\,$s$/$mm$^2$) and ODFs are generated with Generalized Q-Space Imaging (GQI,\citep{Yeh2010,Yeh2011}).} The libraries contain 15,366 (2 fibers, simulations), 46,091 (2 fibers, \textit{in vivo}) and \mrev{R1.1}{4,753,266, (3 fibers \textit{in vivo})} elements and are created in 45$\,$s, 51$\,$s and 6$\,$h$\,$7$\,$min$\,$13$\,$s respectively on a standard high-end laptop \mrev{R1.4}{(Dell Precision 5510, Quad core Intel Xeon E3-1505)}.\par

The ODF-Fingerprinting method is compared to peak identification using local maximum search (DSIStudio\footnote{\url{http://dsi-studio.labsolver.org}} \citep{Yeh2010}, compiled from source on Nov 21st, 2018; a version of the algorithm in Matlab (Mathworks) is also used), Newton search along a set of specified directions (MRtrix3\footnote{\url{https://github.com/MRtrix3}}, v3.0\_RC3, compiled from source on Jan 27th, 2019,\textit{sh2peaks}, \mrev{R1.3}{default parameters}), \mrev{}{multi-shell multi-tissue constrained spherical deconvolution (CSD, MRtrix3, \textit{dwi2fod msmt\_csd}\citep{Jeurissen2014} and \textit{sh2peaks}, default parameters, unsupervised estimation of response functions using \textit{dwi2response dhollander} \citep{Dhollander2016})}and probabilistic estimation (FSL\footnote{\url{https://fsl.fmrib.ox.ac.uk/fsl}} \citep{Jenkinson2012}, \textit{bedpostx} \citep{Behrens2007}, v5.0.9). \mrev{R1.3}{For \textit{bedpostx} the following parameters were used: deconvolution with sticks and a range of diffusivities, constrained non-linear fitting, 250 burn-in points, 500 jumps, automatic relevance determination.}\par

The matching process itself takes 2min14 \mrev{R1.1,R1.2}{(2 fiber library, 1h43min2s for the 3 fiber library)} on a standard high-end laptop \mrev{R1.4}{(Dell Precision 5510, Quad core Intel Xeon E3-1505)} to reconstruct a full-brain 2.5$\,$mm isotropic acquisition with a 46,091-element library \mrev{R1.1,R1.2}{(4,753,266 elements in the 3 fiber library)}. This compares to 12$\,$s, 26$\,$s, 3$\,$min$\,$24$\,$s and 5$\,$h$\,$11$\,$min$\,$25$\,$s for peak finding on the same dataset with local maximum search, Newton search, CSD and probabilistic estimation approaches, respectively.

Source code (Matlab) for the ODF-Fingerprinting approach is available for download at \url{https://bitbucket.org/sbaete/odffingerprinting}.

\subsection{Computer simulations}

Performance of fiber direction identification is evaluated using a composite hindered and restricted diffusion model (CHARMED, \citep{Assaf2005}) as implemented in the Phantomas-software\footnote{\url{http://www.emmanuelcaruyer.com/phantomas.php}} \citep{Caruyer2014ISMRM}. The simulated volumetric fiber configuration was previously used in the HARDI reconstruction challenge at the ISBI 2013 conference. \par

In addition, ODFs of crossing fiber bundles are simulated using an in-house simulation. Diffusion weighted samples are generated as above (Eq. \ref{ODFgen}), with random fiber directions, on a RDSI q-space grid on four shells (250, 1000, 2250, 4000$\,$s$/mm^2$. \mrev{R1.5,R2.1}{Intra-voxel fiber orientation dispersion of 20\degree\citep{Jelescu2017} is added in some simulations and} Rician noise is added where necessary (SNR of $b_0$ indicated). After ODF-reconstruction, fiber directions are identified with ODF-Fingerprinting and the other methods described above.
Results are binned based on the simulated crossing angle to calculate number of identified fibers, detected crossing angular error ($|\alpha_{found} - \alpha_{simulated}|$), angular precision and dispersion of the fiber directions \citep{Kuo2008} and the detected crossing angle. The angular dispersion is a measure of angular accuracy as it assesses the uncertainty of mapping the fiber orientations \citep{Kuo2008}. If more than two fiber directions are found, the directions closest to the simulated directions are used for further calculations.\par


\subsection{In Vivo acquisitions}


In vivo DSI acquisitions of healthy volunteers are acquired on a 3T clinical scanner (Prisma, Siemens, Erlangen; 64ch head coil; 80$\,$mT/m; Twice Refocused Spin Echo EPI sequence; RDSI q-space sampling \citep{Baete2015RDSI}, 4 shells, 200, 1500, 2750, 4000$\,$mm$^2$/s; TR$\,$=$\,$2000$\,$(bootstrap)$\,$/$\,$8500$\,$(whole brain)$\,$ms, TE$\,$=$\,$86$\,$ms, 10$\,$(bootstrap)$\,$/$\,$50$\,$(whole brain) slices, field of view 200$\times$200$\,$mm$^2$, 2.5$\times$2.5$\times$2.5$\,$mm$^3$ resolution, partial Fourier 5/8). For each volunteer either whole brain or limited coverage datasets (five replications) for bootstrapping are acquired. For image correction an extra $b_0$-dataset is acquired with PA phase encoding rather than AP phase encoding. A $T_1$-weighted gradient-echo sequence (MPRAGE) serves as a reference for image registration (TR$\,$=$\,$2300$\,$ms, TE$\,$=$\,$2.87$\,$ms, 192 slices, 1$\times$1$\times$1 mm$^3$ resolution, TI$\,$=$\,$900/1000$\,$ms, 5$\,$min$\,$03$\,$s) The protocol was approved by an Institutional Review Board.\par

\mrev{R2.2}{A high resolution preprocessed \textit{in vivo} DWI acquisition was provided by the Human Connectome Project (HCP) consortium led by Washington University, University of Minnesota, and Oxford University. We used a single subject from the MGH datasets (3T Siemens Skyra System; 64ch head coil; 1000, 3000, 5000$\,$mm$^2$/s, 256 q-space volumes, TR$\,$=$\,$8800$\,$, TE$\,$=$\,$57$\,$ms, 96 slices, field of view 210$\times$210$\,$mm$^2$, 1.5$\times$1.5$\times$1.5$\,$mm$^3$ resolution, partial Fourier 5/8, GRAPPA 3; healthy volunteer).}\par

\subsection{DWI processing}

Post-processing of datasets is performed offline. In-house images are denoised \citep{Veraart2016} and corrected for susceptibility, eddy currents and subject motion using \textit{topup} and \textit{eddy} (FSL \citep{Jenkinson2012}). \mrev{R2.2}{The preprocessed\citep{Glasser2013} HCP images were corrected for gradient non-linearity, motion (FreeSurfer) and eddy currents (FSL\citep{Jenkinson2012} $\textit{eddy}$).} RDSI reconstructions, incorporating variable sample density correction, or GQI\citep{Yeh2010,Yeh2011} reconstructions (HCP dataset) are performed using custom-made software (Matlab, Mathworks) and displayed with DSIStudio \citep{Yeh2010}. Bootstrapped DSI datasets (500) are generated from five original datasets with limited brain coverage using repetition bootknife sampling \citep{Cohen-Adad2011}. From the bootstrapped datasets reproducibility and noise sensitivity metrics are calculated: the number of fibers identified and 95\% confidence intervals (CI), coherence $\kappa$ \citep{Jones2004} and Quantitative Anisotropy (QA) values of the first and second fibers. \par

\mrev{R2.2}{Multi-resolution HCP datasets were created as follows. The 1.5$\,$mm isotropic dataset was down sampled (MRtrix3, $\textit{mrresize}$) to a 3$\,$mm isotropic resolution such that each voxel in the low resolution (LR, 3$\,$mm isotropic) dataset corresponds to 8 high resolution (HR, 1.5$\,$mm isotropic) voxels. Hence, for each 3$\,$mm isotropic voxel we compared the identified fiber directions relative to the fibers found in the 8 corresponding HR voxels. From this comparison we calculated the number of correctly (true positive) and wrongly (false positive) identified fibers and the number of missed fibers (false negative).}

Fiber tracts are generated with a deterministic tracking algorithm (\citep{Yeh2013b}, implemented in DSIStudio \citep{Yeh2010}, parameters as suggested in \citep{Fernandez-Miranda2012}) in both simulated and \textit{in vivo} datasets. Tracts originate from random seeding points uniformly distributed in user-defined seeding regions, propagate along the most prominent fiber direction with a step size of 2$\,$mm, are smoothed with 20\% of the previous direction and halt when the turning angle $>$ 60\degree or QA $<$ 0.25. Resulting fiber tracts shorter than 30$\,$mm and longer than 200$\,$mm are discarded until a predetermined number of fiber tracts ($1\times10^4$) is created or a maximum number of seed points is reached ($1\times10^6$). \nrev{R3.1}{While the choice of these tractography parameters influences the final results, relative differences in the tracts resulting from different peak identification methods remained constant as long as the parameters were consistent for all methods.} \par

\section{Results}

\subsection{Simulation Results}

\begin{figure*}[tbh]
    \begin{center}
    \includegraphics[width=0.75\textwidth,trim=0 0 0 0, clip]{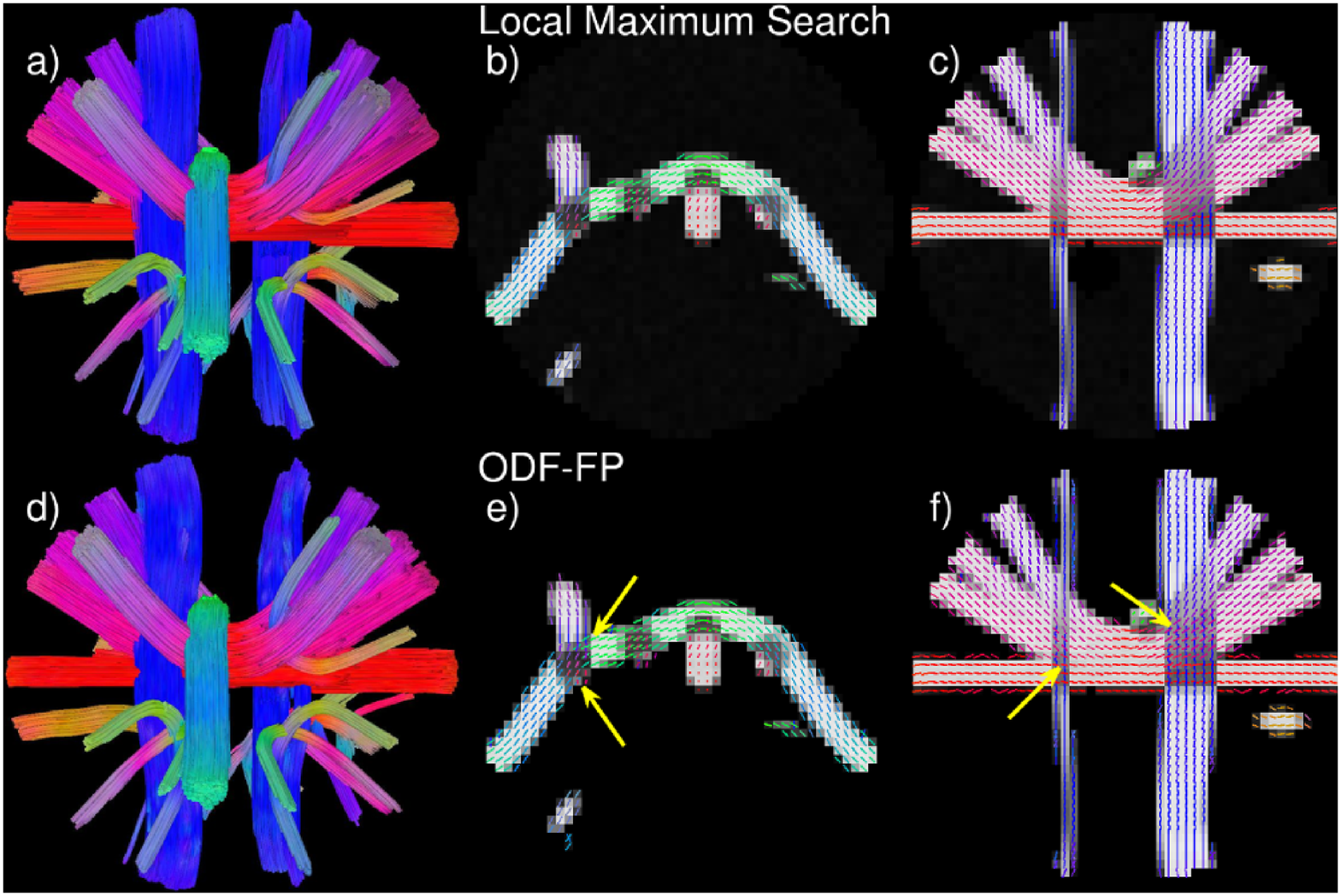}
        \caption{Global fiber tractography (a,d) and identified fiber directions (b,c,e,f) in a dataset simulated with the Phantomas-software. The fiber directions are detected by local maximum search (DSIStudio, top row) and ODF-Fingerprinting (ODF-FP, bottom row). Areas where ODF-FP performed better are indicated with a yellow arrow (e,f).\label{phantomas}}
    \end{center}
\end{figure*}

\begin{figure*}[tbh]
    \begin{center}
    \includegraphics[width=0.75\textwidth,trim=0 0 0 0, clip]{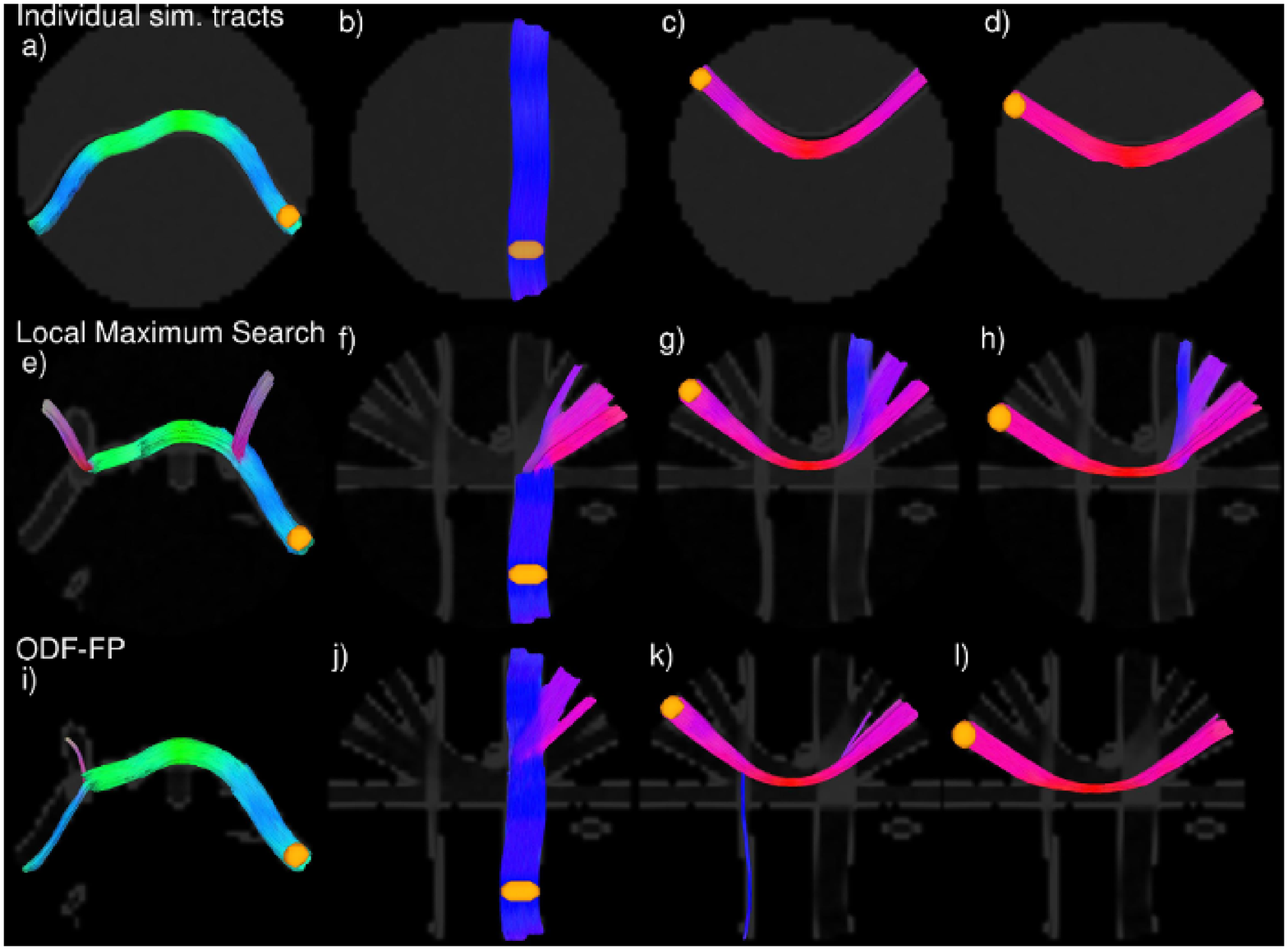}
        \caption{Local fiber tractography of selected fiber tracts (top row) in a dataset simulated with the Phantomas-software using fiber directions detected by local maximum search (DSIStudio, middle row) and ODF-Fingerprinting (ODF-FP, bottom row). The orange ROIs in each image were used to initiate the tractography. \label{phantomastracto}}
    \end{center}
\end{figure*}

Fig. \ref{phantomas} compares the performance of the fiber identification of ODF-FP to the local maximum search algorithm in a diffusion dataset simulated with the Phantomas-software \citep{Caruyer2014ISMRM}. ODF-FP better identifies small crossing angles (Fig. \ref{phantomas}b,c,e,f, yellow arrows) even though the diffusion dataset is simulated with a diffusion model different from the model we used to generate the library (composite hindered and restricted diffusion model \citep{Assaf2005} vs. sum of diffusion tensors \citep{Alexander2001,Tuch2002,Wilkins2015}). \mrev{R2.5}{ODF-FP also correctly identifies more fibers (Fig. {\refsupplphantomas}a,e), produces less false negatives (missed fibers, Fig. {\refsupplphantomas}c,g) and reduces the angular error of identified fiber directions (Fig. {\refsupplphantomas}d,h).} Detection of smaller crossing angles leads to improved fiber tractography in individual simulated fiber bundles (Fig. \ref{phantomastracto}).\par

\begin{figure*}[tbh]
    \begin{center}
    \includegraphics[width=0.95\textwidth,trim=0 0 0 0, clip]{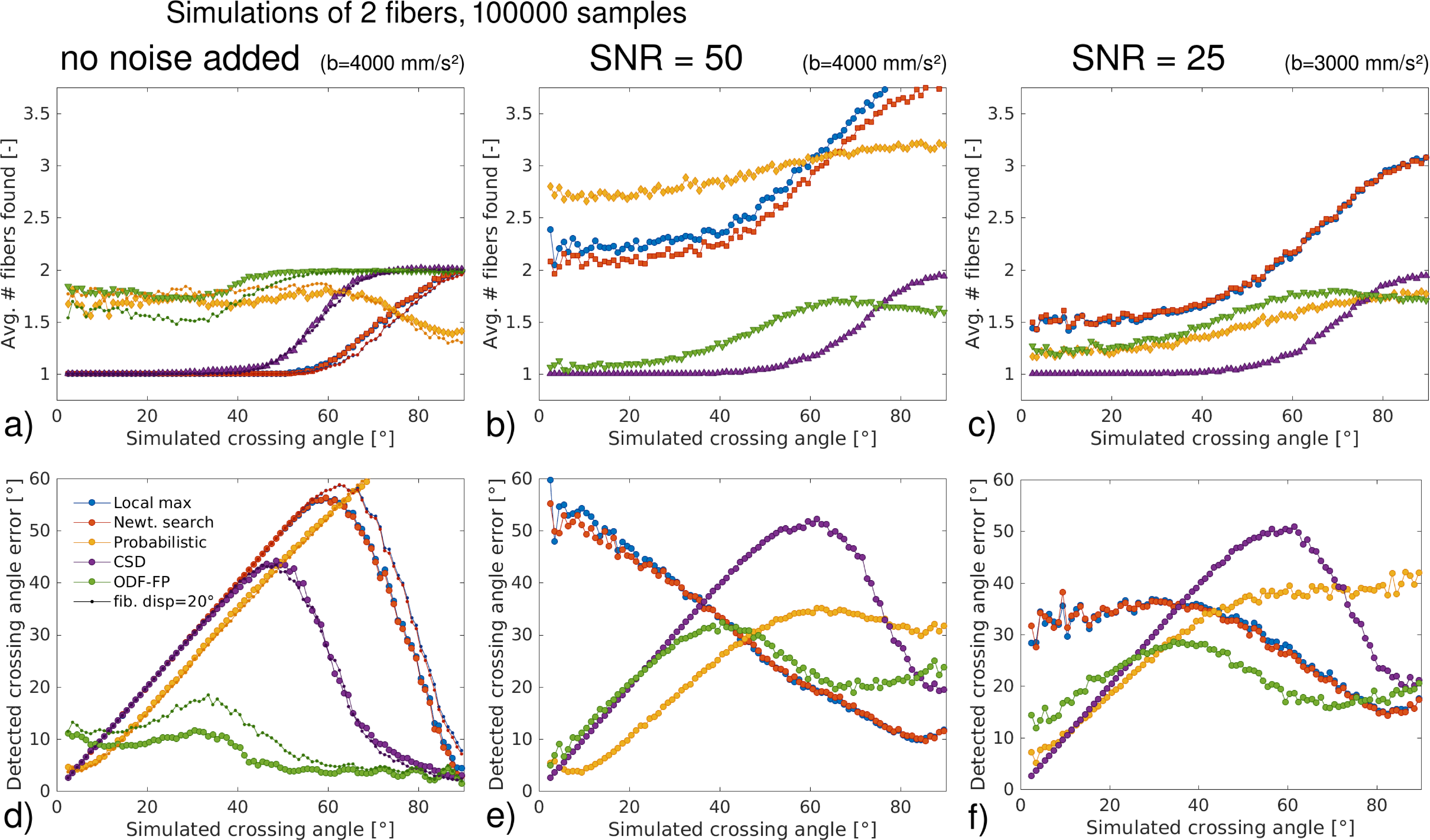}
        \caption{Average number of fibers found (a,b,c) and error on the detected crossing angles (d,e,f) of simulated pairs of crossing fibers (random angle) as identified by local maximum search (DSIStudio), Newton search (MRtrix3), probabilistic estimation (FSL, \textit{bedpostx}), \mrevi{}{constrained spherical deconvolution (MRtrix3, \textit{dwi2fod msmt\_csd})}, and ODF-fingerprinting. The results are plotted as a function of the simulated crossing angle with no added noise (a,d, b$_{max}\,$=$\,$4000$\,$mm/s$^2$), with SNR 50 (b,e, b$_{max}\,$=$\,$4000$\,$mm/s$^2$) and \mrevi{R2.6}{with SNR 25 (c,f, b$_{max}\,$=$\,$3000$\,$mm/s$^2$)}. \mrevi{R1.5,R2.1}{(a) and (d) include  results of simulated pairs of crossing fibers with an intra-voxel fiber orientation dispersion of 20\degree.} \label{odfsim}}
    \end{center}
\end{figure*}

\begin{figure*}[tbh]
    \begin{center}
    \includegraphics[width=0.95\textwidth,trim=0 0 0 0, clip]{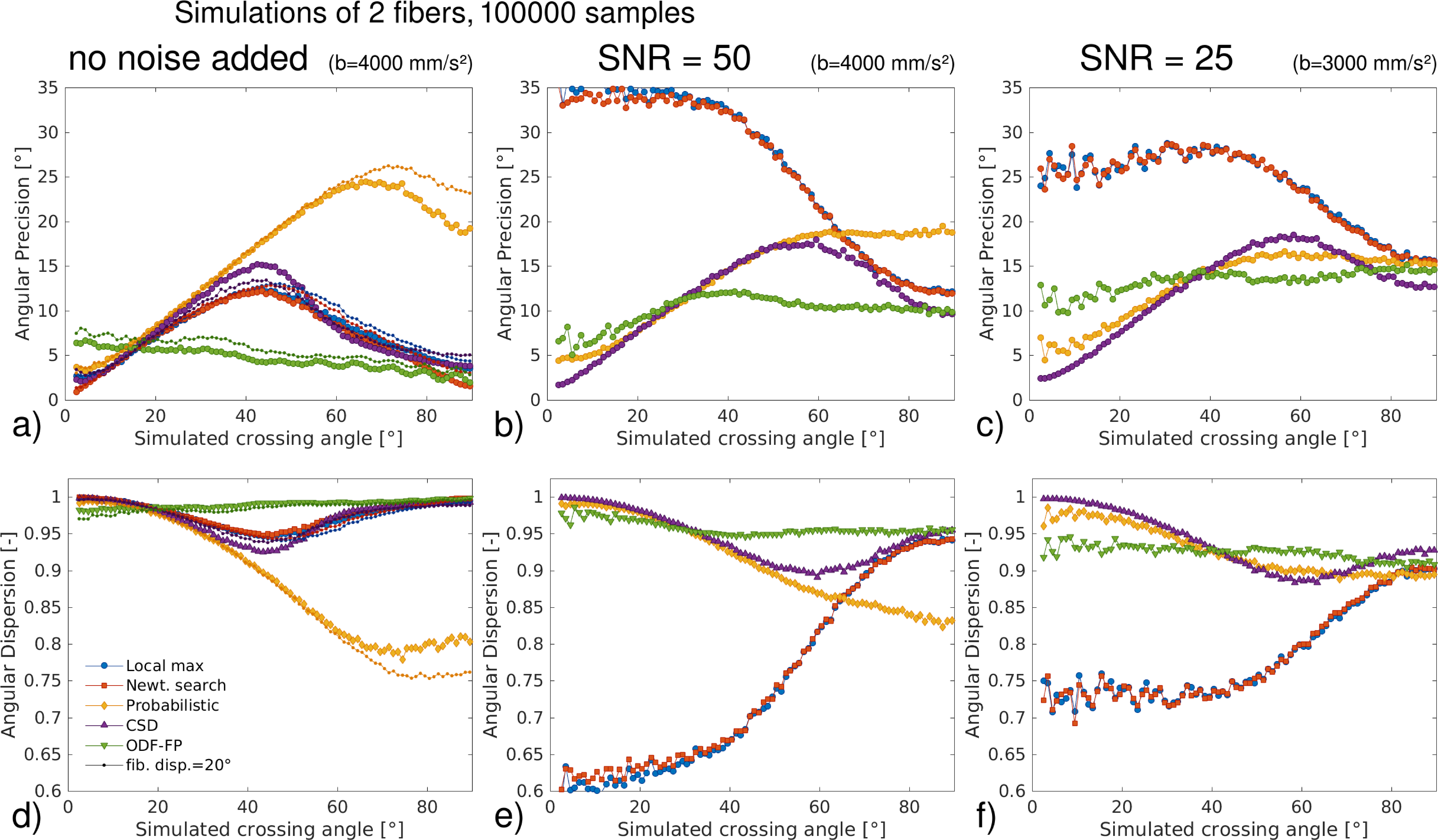}
        \caption{Angular precision (a,b,c) and dispersion (d,e,f) of simulated crossing fibers (random angle) as identified by local maximum search (DSIStudio), Newton search (MRtrix3), probabilistic estimation (FSL, \textit{bedpostx}), \mrevi{}{constrained spherical deconvolution (MRtrix3, \textit{dwi2fod msmt\_csd})}, and ODF-fingerprinting. The results are plotted as a function of the simulated crossing angle with no added noise (a,d, b$_{max}\,$=$\,$4000$\,$mm/s$^2$), with SNR 50 (b,e, b$_{max}\,$=$\,$4000$\,$mm/s$^2$) and \mrevi{R2.6}{with SNR 25 (c,f, b$_{max}\,$=$\,$3000$\,$mm/s$^2$)}. \mrevi{R1.5,R2.1}{(a) and (d) include  results of simulated pairs of crossing fibers with an intra-voxel fiber orientation dispersion of 20\degree.}\label{odfsimprecdisp}}
    \end{center}
\end{figure*}

The results of individual crossing fiber ODF simulations are summarized in Fig. \ref{odfsim}, \ref{odfsimprecdisp} and {\refsupplodfsim}. These simulations, with their inherent reference, allow detailed evaluation of the fiber identification performance as a function of the simulated crossing angle at different SNR levels (no added noise in the left column of Fig. \ref{odfsim}, \ref{odfsimprecdisp} and {\refsupplodfsim}, SNR$\,$=$\,$50 in the middle column \mrev{R2.6}{and SNR$\,$=$\,$25 in the right column). For SNR$\,$=$\,$25, the b-value was limited to 3000$\,$mm/s$^2$ as the SNR of the b$\,$=$\,$4000$\,$mm/s$^2$-shell was too low for processing.} The different search strategies correctly identify most larger crossing angle pairs (i.e. two fibers are identified) when no noise is added but fail clearly when the crossing angles become smaller or noise is added (Fig. \ref{odfsim}a,b,c). In particular, the ODF maximum search methods (local maximum and Newton search) detect too many fiber bundles in noisy data while CSD does not find all fibers; the performance of the probabilistic method seems to depend on the noise level. ODF-FP in contrast successfully finds just two fibers for a range of crossing angles (50$\,$-$\,$90\degree) for noisy data.\par

Looking at the crossing angular error (Fig. \ref{odfsim}d,e,f), the ODF maximum search methods lead to disproportionately large errors when estimating the angle of shallow crossings in the presence of noise. The probabilistic, CSD and ODF-FP methods, on the other hand, show errors that scale with the simulated crossing angle when the crossing angle is small (as only one fiber is identified Fig. \ref{odfsim}b,c). The crossing angular error then drops for larger crossing angles as more fiber pairs are identified. Compared to CSD, ODF-FP is able to detect smaller crossing angle pairs (40\degree$\,$ vs. 50\degree, or 50\degree$\,$ vs. 70\degree$\,$ when the data is noisy; Fig. \ref{odfsim}a-c). In more detail, the ODF maximum search methods and the probabilistic method tend to underestimate large crossing angles (Fig. {\refsupplodfsim}a,b,c), while CSD and ODF-FP mostly correctly estimate these large crossing angles ($>\,$50\degree, Fig. {\refsupplodfsim}a,b,c). At smaller crossing angles ODF-FP, CSD and the probabilistic method do not find all of the crossing fiber pairs. These methods however also do not generate incorrect estimates in noisy datasets in contrast to the ODF maximum search methods (Fig. {\refsupplodfsim}b,c).\par

Fig. \ref{odfsimprecdisp} and {\refsupplodfsim}d-i plot the angular precision and dispersion of the identified fiber directions. While Fig. \ref{odfsimprecdisp} plots the values including all identified fiber combinations, the results in Fig. {\refsupplodfsim}d-i are split by the number of fiber directions found for each ODF ($\Box$ and $\bigcirc$ when two fibers are found vs. $\Diamond$ when one fiber is found). Angular precision and dispersion of the ODF maximum, CSD and ODF-FP search methods is similar when no noise is added (Fig. \ref{odfsimprecdisp}a,d) with ODF-FP showing a higher precision and less dispersion. In general, the angular precision is high at large and small crossing angles when either a single fiber or the average of two crossing fibers is found. The angular precision and dispersion of the probabilistic method is lower (higher value for the precision) since this method does not identify all of the fibers. When noise is added, ODF-FP and CSD retain most of their angular precision and dispersion, while the ODF maxima search methods loose precision due to erroneous detection of fibers (Fig. \ref{odfsimprecdisp}b,c,e,f); the probabilistic method seems to perform better at lower SNR. Angular precision and dispersion split by number of fiber directions (Fig. {\refsupplodfsim}d-i) largely follow the same trend as above.\par

\mrev{R1.5,R2.1}{Intra-voxel fiber orientation dispersion\citep{Jelescu2017} can complicate the identification of fiber directions. Fig. \ref{odfsim}a,d and \ref{odfsimprecdisp}a,d explore the implications of a 20\degree\citep{Jelescu2017} fiber dispersion (smaller dot symbols). These figures show a general decrease in precision and increase in crossing angular error of all methods. Overall, the impact of fiber dispersion is minimal.}

\subsection{In Vivo Results}

\begin{figure*}[tbh]
    \begin{center}
    \includegraphics[width=0.75\textwidth,trim=0 0 0 0, clip]{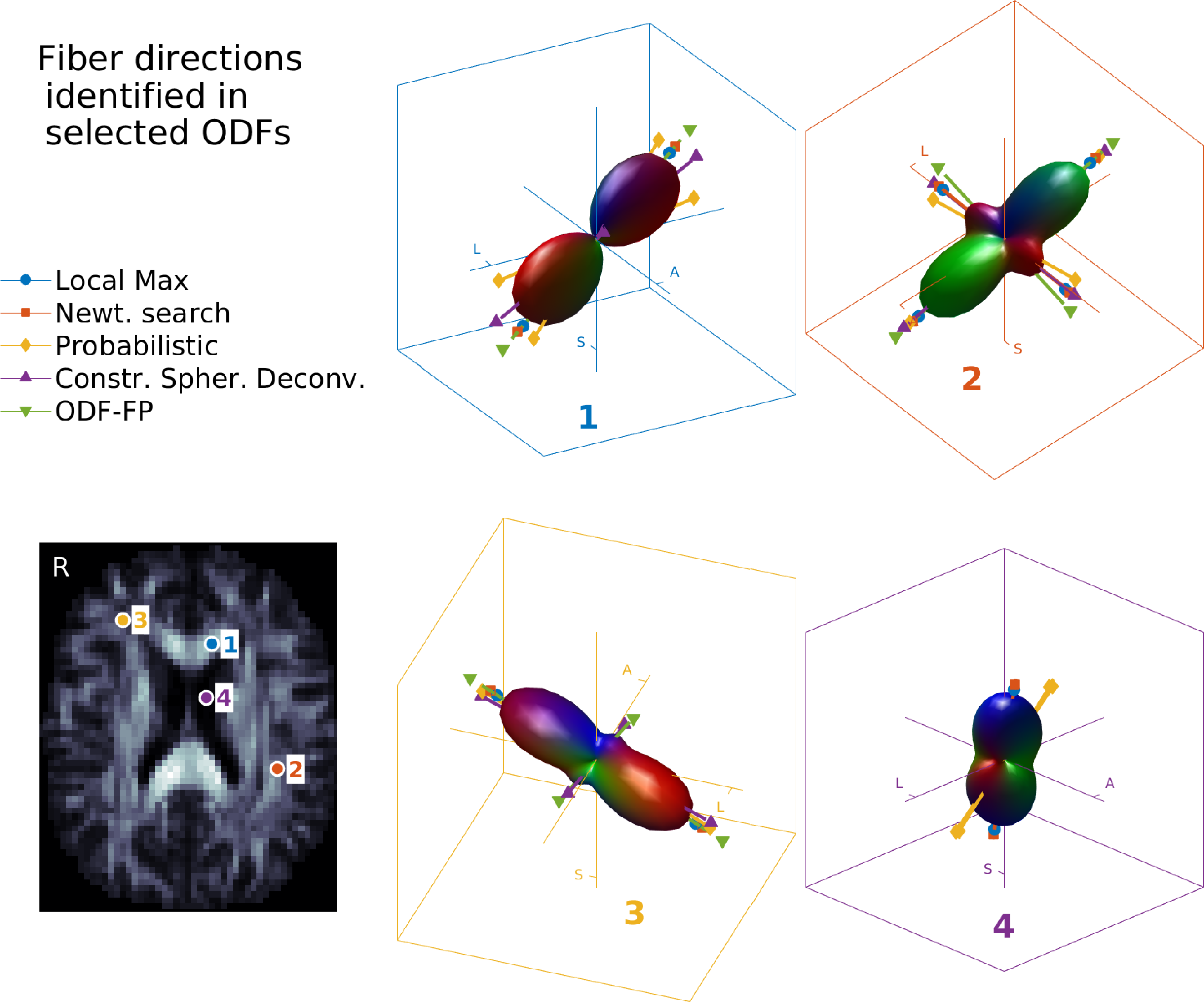}
        \caption{\textit{In vivo} ODFs and fiber directions identified from the ODFs by local maximum search (DSIStudio), Newton search (MRtrix3), probabilistic estimation (FSL, \textit{bedpostx}), \mrevi{}{constrained spherical deconvolution (MRtrix3, \textit{dwi2fod msmt\_csd}}, and ODF-Fingerprinting (ODF-FP) in voxels selected from an RDSI acquisition. ODFs are rotated for visibility, their position is indicated in the bottom left QA-map. \label{bootzoomodf}}
    \end{center}
\end{figure*}

A closer look at a few selected \textit{in vivo} ODFs (Fig. \ref{bootzoomodf}) illustrates the differences in fiber directions identified by the different algorithms. ODF maximum search methods reliably detect strong peaks in the ODFs, but tend to miss more subtle smaller peaks (Fig. \ref{bootzoomodf}, ODF 3). Probabilistic estimation, on the other hand, seems to read too much into minor ODF peaks (Fig. \ref{bootzoomodf}, ODF 1,4). In contrast, ODF-FP and CSD identify the desired peaks whilst refraining from detecting non-existing fiber bundles (Fig. \ref{bootzoomodf}, ODF 3).\par

\begin{figure*}[tbh]
    \begin{center}
    \includegraphics[width=0.9\textwidth,trim=0 0 0 0, clip]{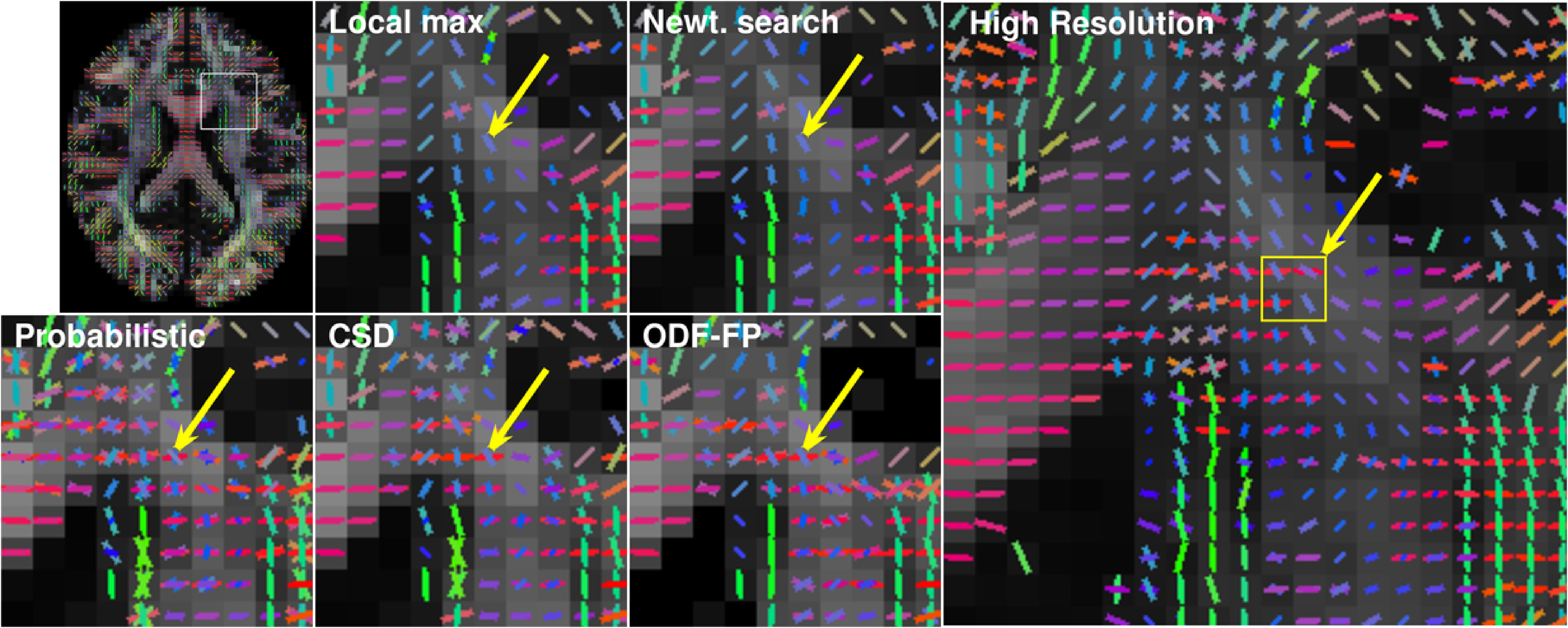}
        \caption{ \mrevi{R2.2}{Fiber directions identified in a subsection of a transversal slice in both high and low resolution versions of a HCP dataset. In the low resolution datasets, the fibers are identified by 5 algorithms: local maximum search (DSIStudio), Newton search (MRtrix3), probabilistic estimation (FSL, \textit{bedpostx}), constrained spherical deconvolution (MRtrix3, \textit{dwi2fod msmt\_csd}, and ODF-Fingerprinting (ODF-FP). The arrows indicate one of the voxels where ODF-FP identified fiber directions which are missed by some of the other algorithms.}\label{DirImages_HCP}}
    \end{center}
\end{figure*}

\begin{figure*}[tbh]
    \begin{center}
    \includegraphics[width=0.75\textwidth,trim=0 0 0 0, clip]{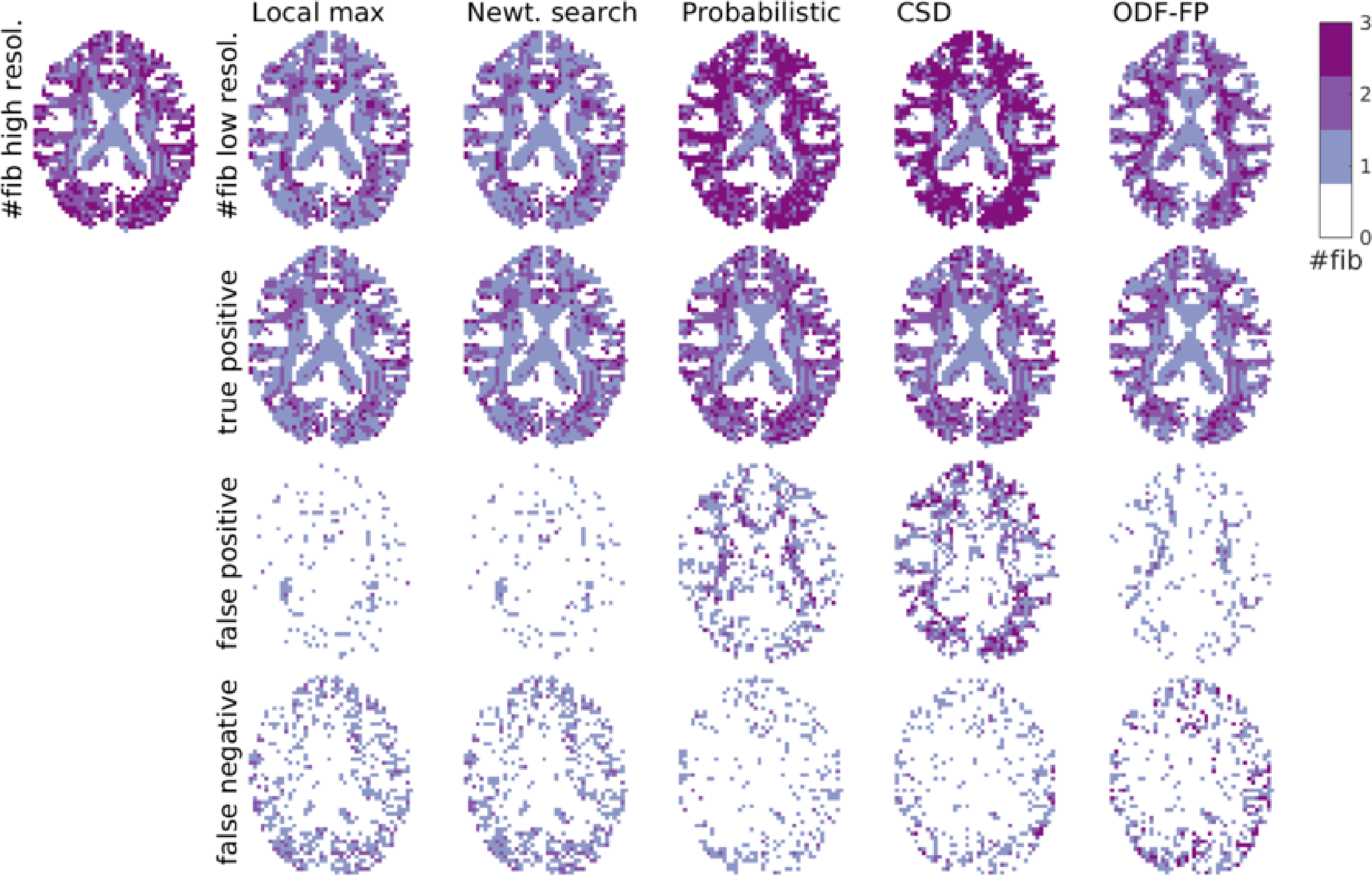}
        \caption{ \mrevi{R2.2}{Maps of the number of fibers identified in the high and low resolution versions of a HCP dataset (top row). The low resolution dataset was processed with 5 algorithms (local maximum search (DSIStudio), Newton search (MRtrix3), probabilistic estimation (FSL, \textit{bedpostx}), constrained spherical deconvolution (MRtrix3, \textit{dwi2fod msmt\_csd}, and ODF-Fingerprinting (ODF-FP) and the number of correctly identified (true positive, second row), wrongly identified (false positive, third row) and missed (false negative, bottom row) fibers were calculated relative to the reference high resolution dataset. ODF-FP shows a larger number of true positive fibers at the cost of a somewhat higher number of false positive fibers.}\label{TruePos_HCP}}
    \end{center}
\end{figure*}

\mrev{R2.2}{Evaluation of \textit{in vivo} fiber direction identification can be performed by comparing fiber directions found in down sampled DWI with those present in high resolution DWI. The latter thus form an internal reference which is otherwise absent in \textit{in vivo} data. Fig. \ref{DirImages_HCP} illustrates the setup of such an experiment; the directions found in a LR voxel (yellow arrows) are compared to those found in the corresponding HR voxels (yellow square in the panel on the right). In the indicated voxel, ODF-FP, CSD and the probabilistic method successfully identify the second fiber bundle. However, the probabilistic method and CSD also identify a number of false positive fibers. Maps of the number of correctly (true positive) and wrongly (false positive) identified fibers (Fig. \ref{TruePos_HCP}, second and third row) confirm these findings over the whole volume.}\par

\begin{figure*}[tbh]
    \begin{center}
    \includegraphics[width=0.95\textwidth,trim=0 0 0 0, clip]{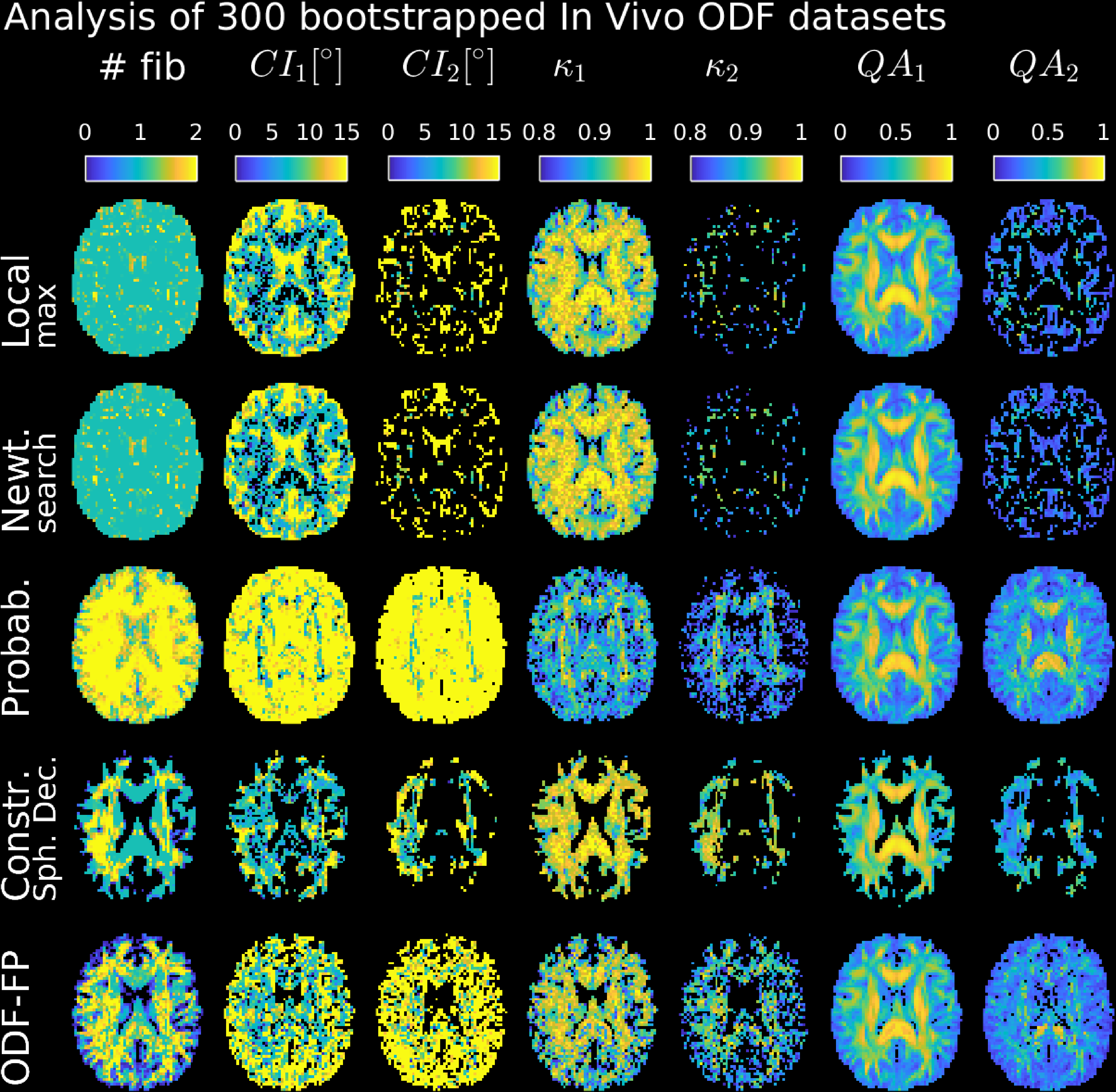}
        \caption{Reproducibility and noise sensitivity analysis of fiber identification in 300 bootstrapped RDSI datasets. Fibers are identified by local maximum search (DSIStudio, top row), Newton search (MRtrix3, 2nd row), probabilistic estimation (FSL, \textit{bedpostx}, 3rd row), \mrevi{}{constrained spherical deconvolution (MRtrix3, \textit{dwi2fod msmt\_csd}, 4th row)}, and ODF-Fingerprinting (ODF-FP, bottom row). Displayed are the number of fibers identified and 95\% confidence intervals (CI), coherence ($\kappa$) and Quantitative Anisotropy (QA) values for the first and second fiber. \label{odfboot}}
    \end{center}
\end{figure*}

Analysis of 300 bootstrapped RDSI datasets (Fig. \ref{odfboot}) shows that ODF-FP identifies crossing fibers where expected (Fig. \ref{odfboot}, \#fib-column on the left) while CSD finds them to a lesser extent, ODF maximum search methods miss fibers and probabilistic estimation identifies an artificially high number of fibers. In addition, CSD and ODF-FP do not identify as many fibers in areas where no fibers are expected such as the Cerebrospinal fluid (CSF). This leads to different patterns in the reproducibility and noise sensitivity statistics (95\% confidence intervals (CI) and coherence $\kappa$, Fig. \ref{odfboot}). QA-maps further illustrate these altered patterns (Fig. \ref{odfboot}, $QA$-columns). \par

When fibers are identified with ODF-FP, the 95\% CI and $\kappa$ of the fibers are similar to the reproducibility values found for ODF maximum search methods and CSD, even for the second identified fiber ($CI_2$, $\kappa_2$-columns of Fig. \ref{odfboot}). The 95\% CI and $\kappa$ maps further show the noise sensitivity of the probabilistic method with higher 95\% CI and lower $\kappa$-values.\par

\begin{figure*}[tbh]
    \begin{center}
    \includegraphics[width=0.95\textwidth,trim=0 0 0 0, clip]{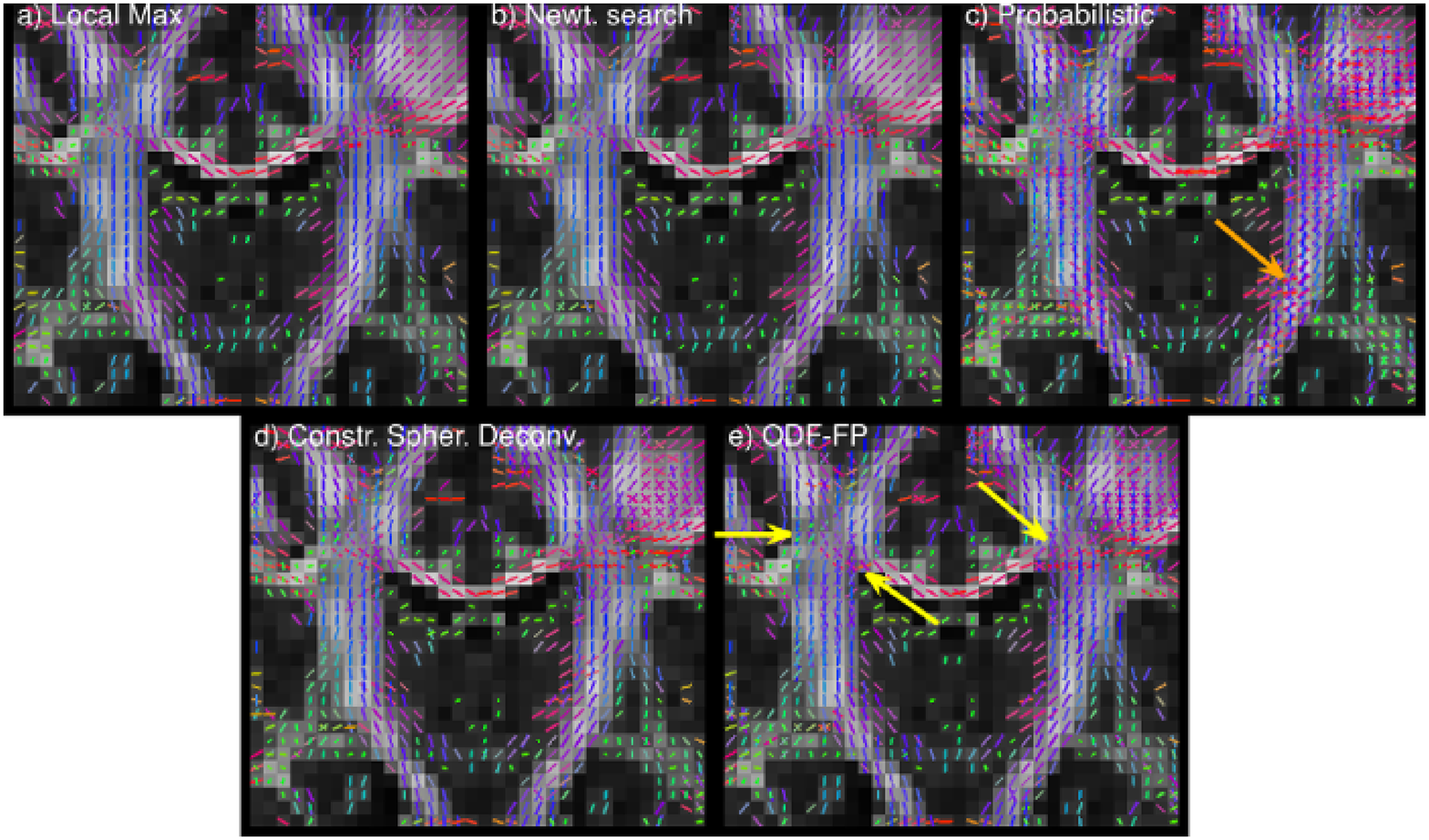}
        \caption{Fiber directions are identified by local maximum search (DSIStudio, a), Newton search (MRtrix3, \textit{shpeaks}, b), probabilistic estimation (FSL, \textit{bedpostx}, c), \mrevi{}{constrained spherical deconvolution (MRtrix3, \textit{dwi2fod msmt\_csd}, d)} and ODF-Fingerprinting (ODF-FP, e) in a coronal slice of a whole brain \textit{in vivo} dataset. Example areas where ODF-FP performed better are indicated with a yellow arrow; an example area where spurious fibers are detected are indicated with an orange arrow.\label{invivodirections}}
    \end{center}
\end{figure*}

\begin{figure*}[tbh]
    \begin{center}
    \includegraphics[width=0.8\textwidth,trim=0 0 0 0, clip]{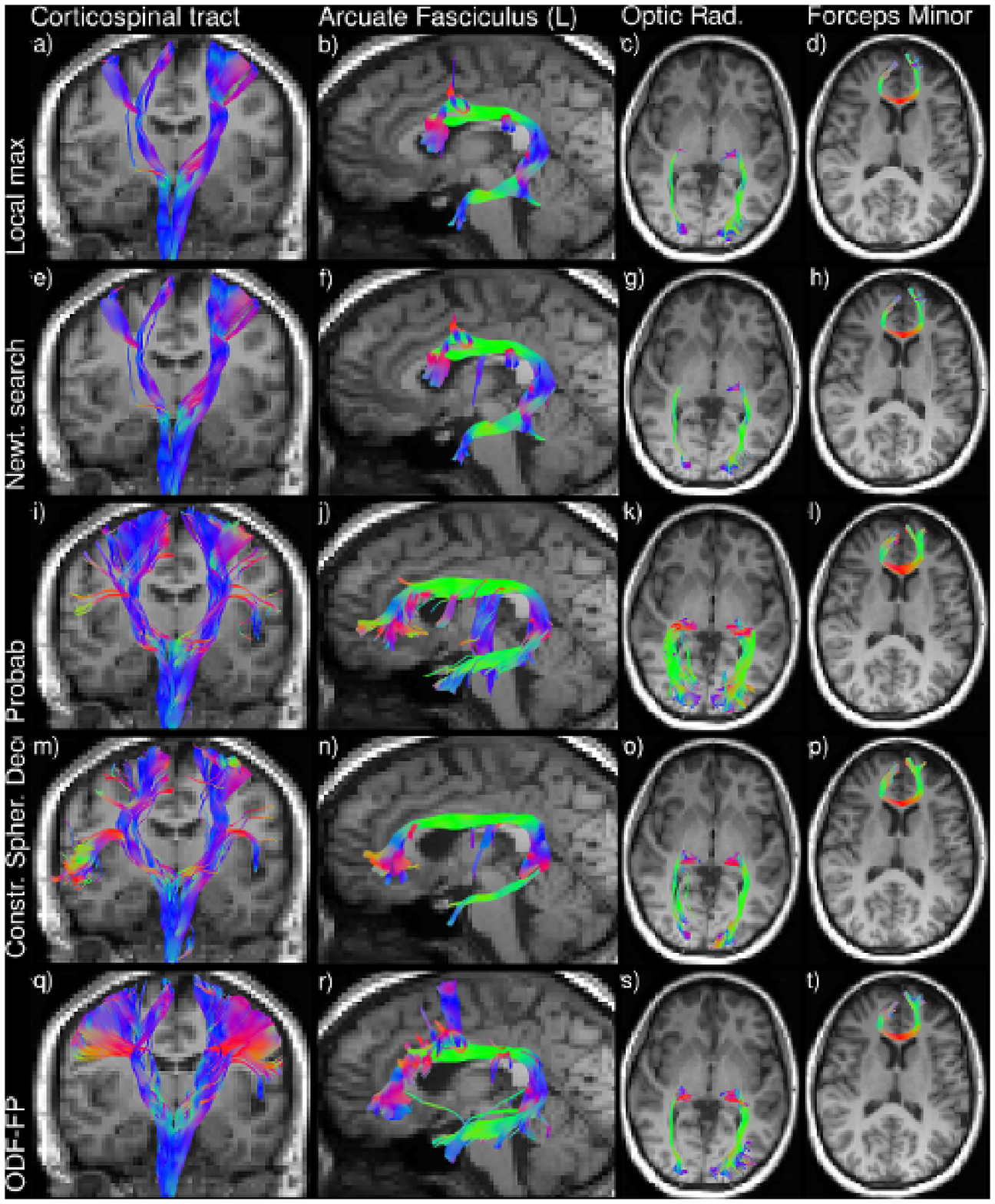}
        \caption{Fiber tractography of the corticospinal tracts (a,e,i,m,q), the left arcuate fasciculus (b,f,j,n,r), the optic radiations (c,g,k,o,s) and the forceps minor (d,h,l,p,t) in a whole brain RDSI dataset. Fiber directions are identified by local maximum search (DSIStudio, a-d), Newton search (MRtrix3, \textit{shpeaks}, e-h), probabilistic estimation (FSL, \textit{bedpostx}, i-l), \mrevi{}{constrained spherical deconvolution (MRtrix3, \textit{dwi2fod msmt\_csd}, m-p)} and ODF-Fingerprinting (ODF-FP, q-t).\label{invivotracto}}
    \end{center}
\end{figure*}

Fiber tractography on a whole brain \textit{in vivo} dataset shows that the fiber directions identified by ODF-FP (Fig. \ref{invivodirections}) allow the algorithm to improve results (Fig. \ref{invivotracto}). Fiber bundles generated with ODF-FP input probe the expected anatomical extent of the tracts in contrast to the fiber bundles based on ODF maximum search methods and CSD, in particular for the Corticospinal tract. The tendency of probabilistic methods to derive erroneous fiber directions (e.g. Fig. \ref{bootzoomodf}, ODF 1,4) produces more spurious tracts (Corticospinal tract, Arcuate Fasciculus and Optic Radiation in particular).\par

\section{Discussion}
As tractography based representations of the brain gain importance in clinical and neuroscientific applications, so grows the desire to resolve evermore detailed and complex neuronal pathways. One of the last remaining methodological challenges on the path towards such high-fidelity tractography representations is the identification of multiple intra-voxel fiber crossings. In particular, when dealing with small crossing angles. Although great progress has already been made \citep{Baete2015RDSI,Descoteaux2007,Frey2008,Berman2008,Tournier2004,Aganj2010}, these methods still fail to reliably detect crossing angles less than 40\degree \citep{Kuo2008,Jeurissen2012,Tournier2008,Descoteaux2009,Daducci2014,Wilkins2015}. In this work, we have sought to overcome these limitations by changing the paradigm for fiber direction identification from a search for maxima on the surface of the ODF to an assessment of similarity relative to a dictionary of pregenerated ODFs.\par

The application of key concepts from fingerprinting to the ODF based fiber direction identification task (Fig. \ref{odffpscheme}e) improves the detection of fiber pairs crossing at small angles as shown in simulation results (Fig. \ref{phantomas},\ref{odfsim}). This is achieved while maintaining angular precision of fiber directions over the whole range of crossing angles (Fig. \ref{odfsimprecdisp}a,b,c {\refsupplodfsim}d,e,f). \textit{In vivo} bootstrap analysis shows that ODF-FP detects crossing fiber pairs where anatomically expected (Fig. \ref{odfboot}) while not over-identifying fiber bundles in areas where no fiber bundles are expected such as in the CSF (Fig. \ref{odfboot}, \# fibers). Furthermore, the fiber directions identified with ODF-FP are reproducible over the bootstrapped datasets (Fig. \ref{odfboot}, CI and $\kappa$) and \mrev{R2.2,R2.3}{reproduce the internal reference of an \textit{in vivo} multi-resolution HCP dataset (Fig. \ref{DirImages_HCP}-\ref{TruePos_HCP})}. Consequently, the improved fiber detection results in increased adherence of fiber tractography to the underlying simulated microstructure (Fig. \ref{phantomastracto}). Although no gold standard is available for \textit{in vivo} tractography, the data suggest that the fingerprinting based method \mrev{R2.3}{to improve tractography results} (Fig. \ref{invivodirections}, \ref{invivotracto}).\par

The performance of the ODF-FP methods compares favorably to ODF maximum search methods (local maximum and Newton search), CSD and probabilistic methods. These methods generally underestimate the crossing angle (Fig. {\refsupplodfsim}a) or, at smaller angles, do not detect the crossing fiber pair but rather a single fiber (Fig. \ref{odfsim}a). While the former situation is suboptimal, the latter biases the tractography (Fig. \ref{phantomastracto}). Both problems originate in the focus on the maxima on the ODF surface in combination with the intrinsic ODF peak width \citep{Barnett2009,Jensen2016}. The ODF maximum search methods perform better when attempting to identify fiber pairs with crossing angles between 40\degree$\,$ and 55\degree$\,$ in noisy data (Fig. \ref{odfsim}e,f), though still overestimate the crossing angle (Fig. {\refsupplodfsim}b,c). This improved detection does come at the cost of erroneous detection of non-existent fiber bundles (Fig. \ref{odfsim}b,c) which might confuse tractography algorithms. The ODF-FP algorithm more precisely determines (smaller) crossing angles. Nevertheless, no approach to fiber identification from ODFs is perfect.\par

The ODF-library in this work is generated with the simple diffusion model described in Eq. \ref{ODFgen}. Results are consistent when this library is applied to simulations using the same generative diffusion model (Fig. \ref{odfsim}, \ref{odfsimprecdisp}, {\refsupplodfsim}), to ODFs simulated with a different diffusion model (CHARMED, Phantomas, Fig. \ref{phantomas}, \ref{phantomastracto}) and to \textit{in vivo} multi-shell HCP (Fig. \ref{DirImages_HCP},\ref{TruePos_HCP}) and RDSI datasets (Fig. \ref{bootzoomodf}, \ref{odfboot}, \ref{invivodirections} and \ref{invivotracto}). \mrev{R2.2}{Results are also consistent when multi-shell sampling and GQI ODF-reconstructions are used (Fig. \ref{DirImages_HCP}-\ref{TruePos_HCP})) rather than RDSI sampling and reconstruction (Fig. \ref{phantomas}-\ref{bootzoomodf},\ref{odfboot}-\ref{invivotracto}). This illustrates that the ODF-Fingerprinting approach can be used regardless of q-space sampling and ODF-reconstruction method.}\par

The diffusion phenomenon \textit{in vivo} is more complex than described by Eq. \ref{ODFgen} \citep{Novikov2016}. Careful consideration of non-Gaussian diffusion in intra- and extra-axonal space, the axon diameter and \mrev{R2.1}{dispersion \cite{Ghosh2016}} of the fiber bundles leads to models such as CHARMED \citep{Assaf2005}, ActiveAx \citep{Alexander2010}, Neurite Orientation Dispersion and Density Imaging \citep{Zhang2012NODDI}, and White Matter Tract Integrity \citep{Fieremans2011}. Each of these models, with an appropriate range of tissue parameters, can be used to generate more complex dictionaries for the ODF-fingerprinting method\mrev{R1.5}{, including fiber configurations such as branching and kissing fibers. By adhering closer to \textit{in vivo} microstructure, these complex dictionaries} may eventually be better suited for ODF-FP and additional model parameters associated with the diffusion models may be simultaneously estimated. By removing the need for direct parameter fitting, we then adhere to the philosophy behind quantitative MR fingerprinting \citep{Ma2013,Cloos2016}.

The computational bottleneck in the ODF fingerprinting method is the matching algorithm (Eq. \ref{dotproductpenalty}). Fortunately, the central dot-product in Eq. \ref{dotproductpenalty}, $L_{ODF}\cdot ODF_m^T$, can be efficiently implemented as a matrix multiplication. Another consideration is the dictionary size. Simulating two crossing fibers with random orientations on a 642 point grid (321 options due to symmetry) gives 321x320$\,$=$\,$102,720 possibilities. This number has to be multiplied with the desired number of fiber parameter combinations. In the ODF-library used here for the \textit{in vivo} reconstructions for instance, 144 parameter combinations were considered leading to a total of $~14.8\times10^6$ possibilities. A major reduction in dictionary size is possible by rotating the maximum value of the ODF-traces to e.g. the Z-axis before matching, reducing the library size by a factor of 321
to $~46.1\times10^3$ in our example for 2 fibers. Further reductions may be possible by rotating the ODF to align the second maximum when present. The ODF-FP approach as currently implemented is slower than the ODF maximum search methods (\mrev{R1.1,R1.2}{2$\,$min$\,$14 for 2 fiber and 1$\,$h$\,$43$\,$min$\,$2$\,$s for 3 fiber dictionaries} compared to 12$\,$s and 26$\,$s for local maximum search and Newton search in a full-brain acquisition); though fast enough for use in general processing pipelines. Acceleration is possible by porting the implementation from Matlab to more general-purpose programming languages. Furthermore, since the ODF-fingerprinting method operates on ODFs, it is easily incorporated in existing postprocessing pipelines.\par

Tractography algorithms guided by the more accurate fiber detection of ODF-FP adhere better to the underlying tissue microstructure (Fig. \ref{invivotracto}), thus enhancing the utility of tractography representations in clinical and neuroscientific applications. In neurosurgery, precise fiber bundle delineation informs surgical decisions \citep{Fernandez-Miranda2012,Shin2012}, while structural brain connectivity analysis is leveraged in neuroscientific research \citep{Jbabdi2015,Galantucci2016,Mitra2016}. Improved tractography will thus aid these endeavors.\par

A limitation of the ODF-FP method is the assumption of a diffusion model. This diffusion model, independent of which model was selected, may be insufficiently generalizable to encompass diseased tissue. Also, in the ODF-library, depending on the model used and number of grid points, the entries may be non-unique. This is expected to be more of a problem when working with more complex diffusion models with multiple parameters. A more complete validation of the ODF-FP and other fiber direction identification methods can be performed using hollow fiber phantoms \citep{Guise2016,Hubbard2015,Pathak2017ISMRM3454}.\par

In conclusion, we demonstrated a novel method for fiber direction identification from ODFs based on key concepts first introduced in MR Fingerprinting. In ODF-Fingerprinting, fiber configurations are selected based on the similarity of the shape of measured ODFs with pre-computed library elements. This approach improves detection of fiber pairs with small crossing angles while maintaining fiber direction precision. The resulting, more precise, fiber directions aid fiber tracking algorithms in more accurately calculating brain connectivity for clinical and neuroscientific applications.\par

\section{Acknowledgements}
This project is supported in part by the National Institutes of Health (NIH, R01\-CA111996, R01\-NS082436 and R01\-MH00380). \mrev{R2.2}{Some of the data were provided by the Human Connectome Project, WU-Minn Consortium (Principal Investigators: David Van Essen and Kamil Ugurbil; 1U54MH091657) funded by the 16 NIH Institutes and Centers that support the NIH Blueprint for Neuroscience Research; and by the McDonnell Center for Systems Neuroscience at Washington University.}

\section{References}

\bibliography{dsi}


\end{document}